\documentclass[12pt,reqno]{amsart}

\usepackage{amsmath,
amsthm,
cite}

\usepackage[footnotesize,hang,sf]{caption}
\usepackage{fullpage}

\usepackage{times,txfonts}

\usepackage{psfrag,color}
\usepackage[dvips]{graphicx}
\numberwithin{equation}{section}
\newtheorem{theorem}{Theorem}
\newtheorem{prop}[theorem]{Proposition}

\DeclareMathOperator{\Imaginary}{Im}
\DeclareMathOperator{\Real}{Re}

\DeclareMathOperator{\Li}{Li_2}

\DeclareMathOperator{\Vol}{Vol}

\usepackage{titlesec}
\titleformat{\section}
{\bfseries\scshape\centering}
{\thesection.}{.5em}{}
\titleformat{\subsection}
{\rmfamily\bfseries}
{\thesubsection}{.5em}{}

\begin{document}
\baselineskip 17pt

\renewcommand{\thefootnote}{\fnsymbol{footnote}}

\sloppy




\title[The Colored Jones Polynomial and the A-Polynomial]{
  Asymptotics  of the Colored Jones Polynomial and the A-Polynomial
}


    \author{Kazuhiro \textsc{Hikami}}


  \address{Department of Physics, Graduate School of Science,
    University of Tokyo,
    Hongo 7--3--1, Bunkyo, Tokyo 113--0033, Japan.
    }

    \urladdr{http://gogh.phys.s.u-tokyo.ac.jp/{\textasciitilde}hikami/}

    \email{\texttt{hikami@phys.s.u-tokyo.ac.jp}}


\vspace{18pt}
\date{July 21, 2004}

\begin{abstract}
We  reveal a relationship between  the colored Jones polynomial
and the A-polynomial  for
twist knots.
We  demonstrate that an asymptotics of the $N$-colored Jones
polynomial in large $N$ gives the potential function, and that
the A-polynomial can be computed.
We also discuss on a case of torus knots.

\end{abstract}



\subjclass[2000]{
}


\maketitle
\section{Introduction}

The $N$-colored Jones polynomial 
$J_{\mathcal{K}}(N)$
is a quantum invariant which is
defined based on  the  $N$-dimensional irreducible  representation of 
the quantum group $U_q(sl(2))$.
Motivated by \emph{Volume Conjecture}
raised by Kashaev~\cite{Kasha96b},
it was pointed out
that the colored Jones polynomial at a specific value
should be  related
to the hyperbolic volume of knot complement~\cite{MuraMura99a}.

As another example of the knot invariant related to
$SL(2; \mathbb{C})$,
we have the
A-polynomial~\cite{CCGLS94a,CoopLong98a}.
This is defined as  an algebraic curve  of eigenvalues of the
$SL(2; \mathbb{C})$ representation of the boundary torus of knot,
and contrary to the quantum invariants such as the colored Jones polynomial
it includes many geometrical informations such as the boundary slopes
of the knot.

Those two knot invariants are superficially independent.
Though, it is recently conjectured~\cite{SGarou03b}
that
the homogeneous  difference equation of the $N$-colored Jones
polynomial for knot $\mathcal{K}$
with respect to $N$ gives the A-polynomial for $\mathcal{K}$
(\emph{AJ conjecture}).
This fact was originally verified
for both the trefoil and the figure-eight knot
with a help of computer algebraic system~\cite{SGarou03b}, and
was later proved for the torus knots~\cite{KHikami04a}.
We should note that  in Ref.~\citen{TTakata04a}
a recursion relation of  the summand  of the colored
Jones polynomial  for the twist knots
was shown to give the A-polynomial.
See also Refs.~\citen{RGelc02a,GelcSain03a}.

Recently pointed out is still
another connection between the colored
Jones polynomial and the A-polynomial.
It was  demonstrated~\cite{SGuko03a,MuraYoko04a} that 
the A-polynomial has a relationship with
an asymptotic limit of  the colored Jones polynomial for a case of the
figure-eight knot.
Our purpose in this article
is to  show that this correspondence is also supported
for a case of the twist knots and the torus knots.

We recall  a fact~\cite{GaroTQLe03a}
that
the $N$-colored Jones function 
$J_\mathcal{K}(N)$ for knot $\mathcal{K}$
can be written in a form of the
$q$-hypergeometric function.
Once we obtain an invariant in the form of the $q$-hypergeometric series,
we may  define the $H$-function~\cite{HMura02a} for knot $\mathcal{K}$
based on  the integrand of an asymptotics of the
$N$-colored Jones polynomial for knot $\mathcal{K}$ as
\begin{equation}
  \label{Jones_and_integrand}
  J_\mathcal{K}(N)
  \sim
  \iiint \mathrm{d} \boldsymbol{x} \,
  \exp
  \left(
    \frac{N}{2 \, \pi \, \mathrm{i} \, r}    \,
    H_\mathcal{K}(\boldsymbol{x} , m^2)
  \right)
\end{equation}
in a limit
\begin{align}
  N & \to \infty
  &
  r & = \text{fixed}
  \label{our_limit}
\end{align}
Here we set a parameter $q$ of the $N$-colored Jones polynomial  as
\begin{equation}
  \label{parameter_q}
  q= \exp \left( \frac{2 \, \pi \, \mathrm{i} \,r}{N} \right)
\end{equation}
and define  a parameter $m$ by
\begin{align}
  \label{define_m}
  m^2
  &= q^N
  \\
  & = \mathrm{e}^{2 \pi \mathrm{i} r}
  \nonumber
\end{align}
In this article we demonstrate
for twist knots
$\mathcal{K}_p$
(Fig.~\ref{fig:twist})
and torus knots $\mathcal{T}_{2,2p+1}$
(Fig.~\ref{fig:torus})
that the $H$-function  is regarded as
the potential function~\cite{NeumZagi85a,TYoshi85a} 
under a  constraint
\begin{subequations}
  \label{H_constraint}
  \begin{gather}
    \label{H_constraint_2}
    x_i \,
    \frac{\partial H_\mathcal{K}(\boldsymbol{x} , m^2)}{
      \partial x_i
    }
    = 0
  \end{gather}
  and that we have
  \begin{gather}
    m^2 \,
    \frac{\partial H_\mathcal{K}(\boldsymbol{x} , m^2)}{
      \partial  (m^2)
    }
    = \log \ell
    \label{H_constraint_1}
  \end{gather}
\end{subequations}
Note that a constraint~\eqref{H_constraint_2}
denotes a  saddle point equation for
the integral~\eqref{Jones_and_integrand} in large $N$ limit.
Eliminating $\boldsymbol{x}$ from a set of
eqs.~\eqref{H_constraint}, we
obtain an algebraic equation of $\ell$ and $m^2$ which coincides with
the A-polynomial of knot $\mathcal{K}$;
the $SL(2; \mathbb{C})$
representation of the meridian $\mu$ and the longitude $\lambda$ of
the boundary torus of knot $\mathcal{K}$
is given by the upper triangular matrices,
\begin{align*}
  \rho(\mu)
  & =
  \begin{pmatrix}
    m & * \\
    0 & m^{-1}
  \end{pmatrix}
  &
  \rho(\lambda)
  & =
  \begin{pmatrix}
    \ell & * \\
    0 & \ell^{-1}
  \end{pmatrix}
\end{align*}
up to conjugation.
This shows~\cite{SGuko03a} an intriguing correspondence between the
color $N$ of the
quantum  knot invariant
and the eigenvalue of the
$SL(2; \mathbb{C})$ representation of the meridian.


This paper is organized as follows.
In Section~\ref{sec:twist} we study the twist knots $\mathcal{K}_p$.
Using the $q$-hypergeometric expression of the colored Jones
polynomial derived in Ref.~\citen{GMasb03a}, we show that the $H$-function with
constraints~\eqref{H_constraint} gives the A-polynomial for the twist
knots which was computed in Ref.~\citen{HosteShana03a}.
We also discuss on a relationship with the volume conjecture, and
study a limit $p\to \pm \infty$.
In Section~\ref{sec:torus}
we show
that this correspondence also works for the torus knot.

Throughout this paper
we use a standard notation~\cite{Andre76};
the $q$-product and
the $q$-binomial coefficient are respectively defined as follows;
\begin{gather*}
  (x)_n
  =
  (x; q)_n
  =
  \prod_{i=1}^n
  \left(
    1 - x \, q^{i-1}
  \right)
  \\[2mm]
  \begin{bmatrix}
    n \\ k
  \end{bmatrix}_q
  =
  \frac{(q)_n}{(q)_{n-k} \, (q)_k}
\end{gather*}

\section{Twist Knot}
\label{sec:twist}

\begin{figure}[htbp]
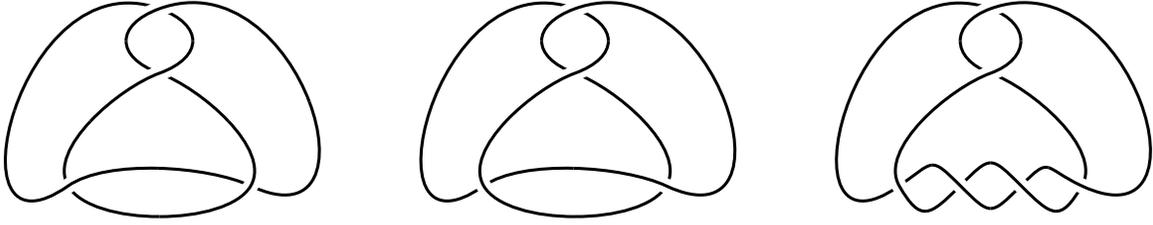

  \centering
  \begin{tabular}{cp{8mm}cp{8mm}c}
    \includegraphics[scale=0.57]{twist_m1.ps}
    & &
    \includegraphics[scale=0.57]{twist1.ps}
    & &
    \includegraphics[scale=0.57]{twist2.ps}
  \end{tabular}
  \caption{Twist knots $\mathcal{K}_p$ are depicted.
    There is a $p$-full twist at the bottom of each figure.
    From left to right,
    figure-eight knot ($p=-1$),
    left-hand trefoil ($p=1$),
    and 
    Stevedore's ribbon knot
    ($p=2$).}
  \label{fig:twist}
\end{figure}

We study the $N$-colored Jones polynomial for the twist knot
$\mathcal{K}_p$.
A case of $p=-1$ is the figure-eight knot
(see Figure~\ref{fig:twist}).

The $N$-colored Jones polynomial for the $p$-twist knot $\mathcal{K}_p$
was computed skein-theoretically in Ref.~\citen{GMasb03a} as follows.
\begin{prop}[\cite{GMasb03a}]
  The $N$-colored Jones polynomial $J_{\mathcal{K}}(N)$ for the twist
  knot
  $\mathcal{K}=\mathcal{K}_p$
  is given as follows (we set $p>0$);
  \begin{multline}
    \label{Jones_twist}
    J_{\mathcal{K}_{p>0}}(N)
    =
    \sum_{s_p \geq \dots \geq s_2 \geq s_1 \geq 0}^\infty
    q^{(p-1) \, s_p  \, (s_p+1) + s_p} \,
    (q^{1-N})_{s_p} \, (q^{1+N})_{s_p}
    \\
    \times
    \left(
      \prod_{i=1}^{p-1}
      q^{s_i^{~2}-(2 s_p +1) s_i} \,
      \begin{bmatrix}
        s_{i+1} \\
        s_i
      \end{bmatrix}_q
    \right) \,
  \end{multline}
  and
  \begin{multline}
    \tag{\ref{Jones_twist}$^\prime$}
    J_{\mathcal{K}_{-p<0}}(N)
    =
    \sum_{s_p \geq \dots \geq s_2 \geq s_1 \geq 0}^\infty
    (-1)^{s_p} \, q^{-(p-\frac{1}{2}) \,  s_p \,  (s_p+1)} \,
    (q^{1-N})_{s_p} \, (q^{1+N})_{s_p}
    \\
    \times
    \left(
      \prod_{i=1}^{p-1}
      q^{-s_i^{~2}+(2 s_p +1) s_i} \,
      \begin{bmatrix}
        s_{i+1} \\
        s_i
      \end{bmatrix}_{q^{-1}}
    \right) 
  \end{multline}
  Here the colored Jones polynomial is normalized such that
  $J_{\text{unknot}}(N)=1$.
\end{prop}

We set a quantum parameter $q$ as in eq.~\eqref{parameter_q},
and study an asymptotic behavior of the quantum invariant in a limit
$N\to\infty$~\eqref{our_limit}.
A limit in
a case of $r=1$,
\emph{i.e.},
$q=\exp( 2 \pi \mathrm{i}/N)$,
corresponds to the ``Volume
Conjecture''~\cite{Kasha96b,MuraMura99a}.

\begin{prop}
  In a  limit~\eqref{our_limit}
  we have
  \begin{equation}
    J_{\mathcal{K}_p}(N)
    \sim
    \iiint \mathrm{d} x_0 \cdots \mathrm{d} x_{p-1}
    \exp \left(
      \frac{N}{2 \, \pi \, \mathrm{i} \, r}
      H_{\mathcal{K}_p}(x_0,\dots,x_{p-1}, m^2)
      \right)
  \end{equation}
  Here $m$ is defined by eq.~\eqref{define_m}, and we have
  \begin{multline}
    \label{define_H}
    H_{\mathcal{K}_{p>0}}(x_0, \dots, x_{p-1}, m^2)
    =
    \sum_{i=1}^{p-1}
    \left(
      \log \left({x_i}/{x_0}\right)
    \right)^2
    +
    \Li(m^2) + \Li(1/m^2)
    \\
    - \Li(x_0/m^2) - \Li(m^2 \, x_0)
    - \Li(x_0)
    +
    \sum_{i=0}^{p-1} \Li(x_{i}/x_{i+1}) - (p-1)\frac{\pi^2}{6}
  \end{multline}
  and  
  \begin{multline}
    \tag{\ref{define_H}$^\prime$}
    \label{define_H_negative}
    H_{\mathcal{K}_{-p<0}}(x_0, \dots, x_{p-1}, m^2)
    =
    - \left( \log x_0 \right)^2
    -\sum_{i=1}^{p-1}
    \left(
      \log \left({x_i}/{x_0}\right)
    \right)^2
    +
    \Li(m^2) + \Li(1/m^2)
    \\
    - \Li(x_0/m^2) - \Li(m^2 \, x_0)
    - \Li(x_0)
    -
    \sum_{i=0}^{p-1} \Li(x_{i+1}/x_i) + (p-1)\frac{\pi^2}{6}
  \end{multline}
  where  we have set  $x_p=1$.
\end{prop}

\begin{proof}
  {}From a definition of the $q$-product,
  we have in a limit $N\to\infty$
  (see \emph{e.g.} Ref.~\citen{RichmSzeke81a,Andre76})
  \begin{equation}
    \log(x \, q)_n
    \sim
    \exp
    \frac{N}{2 \, \pi \, \mathrm{i} \, r}
    \left(
      \Li(x) - \Li(x \, q^n)
    \right)
    \label{log_and_dilog}
  \end{equation}
  Setting
  \begin{align*}
    q^N  & = m^2 
    &
    q^{s_i} & = x_{p-i}
  \end{align*}
  we obtain the $H$-function~\eqref{define_H} and~\eqref{define_H_negative}.
\end{proof}

Our main theorem is as follows.
\begin{theorem}
  \label{prop:previous}
  The function $H_{\mathcal{K}_p}(x_0,\dots,x_{p-1},m^2)$
  defined in eqs.~\eqref{define_H} and~\eqref{define_H_negative}
  is  the potential function~\cite{NeumZagi85a}
  for the
  $p$-twist knot $\mathcal{K}_p$
  under a constraint~\eqref{H_constraint_2}.
  Eliminating $\boldsymbol{x}$ with a condition~\eqref{H_constraint_1}
  gives the A-polynomial of the twist knot $\mathcal{K}_p$.
\end{theorem}



To prove this theorem
we  rewrite
a set of equations~\eqref{H_constraint} for $\mathcal{K}_{p>0}$
as
\begin{subequations}
  \label{saddle_positive}
  \begin{gather}
    \frac{1-m^2 \, x_0}{x_0 - m^2}= \ell ,
    \label{saddle_positive_1}
    \\[2mm]
    \left(1-\frac{x_0}{x_1} \right) \,
    \left(
      \prod_{i=1}^{p-1} x_i^{~2}
    \right)
    =
    x_0^{~2(p-1)} \, (1-x_0) \,
    \left(
      1 - m^2 \, x_0
    \right) \,
    \left(
      1-\frac{x_0}{m^2}
    \right)
    \label{saddle_positive_2}
    \\[2mm]
    x_0^{~2} \,
    \left(
      1- \frac{x_i}{x_{i+1}}
    \right)
    =
    x_i^{~2} \,
    \left(
      1- \frac{x_{i-1}}{x_i}
    \right)
    \qquad \text{for $i=1,\dots,p-1$}
    \label{saddle_positive_3}
  \end{gather}
\end{subequations}
The first equation is solved as
\begin{equation}
  x_0 
  =\frac{1+ \ell \, m^2}{\ell + m^2},
  \label{value_x0}
\end{equation}
When we set
\begin{equation}
  \label{x_and_C}
  x_{p-k}=x_0 \, C_{k}(x_0)
\end{equation}
for $k=1,2,\dots,p-1$,
we see that a rational function $C_k(x)$ is recursively defined by
\begin{gather}
  \label{define_C}
  C_{k+2}(x) = C_{k+1}(x) - \frac{1}{C_{k+1}(x)} + \frac{1}{C_k(x)}
\end{gather}
where the first two rational functions  are given by
\begin{gather*}
  C_0(x) = \frac{1}{x}
  \\[2mm]
  C_1(x) = \frac{1 - (1-x) \,
    \left( 1- m^2 \, x \right) \,
    \left( 1-x/m^2 \right)}{x}
\end{gather*}

In the same way, a set of equations~\eqref{H_constraint} for
a negative case,
$\mathcal{K}_{-p<0}$,
is explicitly written as
\begin{subequations}
  \begin{gather}
    \frac{1 - m^2 \, x_0}{ x_0 - m^2 } = \ell,
    \tag{\ref{saddle_positive_1}}
    \\[2mm]
    \left(
      \prod_{i=1}^{p-1} x_i^{~2}
    \right) \,
    \left(1 - \frac{x_0}{m^2} \right) \,
    \left( 1 - m^2 \, x_0 \right) \,
    (1 - x_0)
    =
    x_0^{~2p} \, 
    \left( 1 - \frac{x_1}{x_0} \right) ,
    \tag{\ref{saddle_positive_2}$^\prime$}
    \label{saddle_negative_2}
    \\[2mm]
    x_0^{~2} \left(1 - \frac{x_i}{x_{i-1}} \right)
    =
    x_i^{~2} \left(1 - \frac{x_{i+1}}{x_{i}} \right),
    \qquad
    \text{for $i=1,\dots,p-1$}
    \tag{\ref{saddle_positive_3}$^\prime$}
    \label{saddle_negative_3}
  \end{gather}
\end{subequations}
In this case $x_0$ is also fixed by eq.~\eqref{value_x0}.
When we also set
\begin{align}
  x_{p-k} & = x_0 \, C_{-k}(x_0)
  \tag{\ref{x_and_C}$^\prime$}
\end{align}
for $k=1, 2, \dots,p-1$,
a rational function $C_{-k}(x)$ is recursively defined from
\begin{gather}
  \tag{\ref{define_C}$^\prime$}
  \frac{C_{-k-2}(x) - C_{-k-1}(x)}{C_{-k-1}(x) - C_{-k}(x)}
  =
  C_{-k-1}(x) \, C_{-k-2}(x)
  \label{define_C_negative}
\end{gather}
where the initial conditions are
\begin{gather*}
  C_0(x)
  =\frac{1}{x}
  \\[2mm]
  C_{-1}(x)
  =
  \frac{m^2 \,x}{
    m^2 \, x^2 - (1-x) (1-m^2 \, x)(m^2-x)
  }
\end{gather*}

As a result,
a set of equations~\eqref{H_constraint} reduces to  an algebraic equation
of $\ell$ and $m$;
\begin{equation}
  C_p(x_0)=1
  \label{f_and_1}
\end{equation}
where the rational function $C_p(x)$ is recursively computed
as above, and
$x_0$ is defined in eq.~\eqref{value_x0}.
Then to prove Theorem~\ref{prop:previous},
we must show that eq.~\eqref{f_and_1} gives the A-polynomial
of the twist knot $\mathcal{K}_p$.

We introduce rational function $\widetilde{C}_p(\ell,m)$ by
\begin{equation}
  \label{define_tilde_C}
  1-C_p(x_0)
  =
  (1-\ell) \,
  ( 1-m^2 ) \,
  \widetilde{C}_p(\ell,m)
\end{equation}
Eqs.~\eqref{define_C} and~\eqref{define_C_negative}
are  transformed into the recursion relation for
$\widetilde{C}_{k}(\ell,m)$;
in a case of positive twist knot
$\mathcal{K}_{p>0}$, we have
\begin{gather}
  \label{recursion_tilde}
  \frac{\widetilde{C}_k - \widetilde{C}_{k+1}}{
    \widetilde{C}_{k+1} - \widetilde{C}_{k+2}
  }
  =
  \left(
    1-
    (1-\ell) \, (1-m^2) \,
    \widetilde{C}_k(\ell,m)
  \right)
  \left(
    1-
    (1-\ell) \, (1-m^2) \,
    \widetilde{C}_{k+1}(\ell,m)
  \right)
\end{gather}
with initial conditions
\begin{align*}
  \widetilde{C}_0(\ell,m)
  & =
  \frac{1}{1+ \ell \, m^2}
  \\[2mm]
  \widetilde{C}_1(\ell,m)
  &=
  \frac{
    \ell+m^6
  }{
    m^2 \, (\ell+m^2 )^2
  }
\end{align*}
and for a negative case
\begin{equation}
  \tag{\ref{recursion_tilde}$^\prime$}
  \label{recursion_tilde_negative}
  \frac{
    \widetilde{C}_{-k-2} - \widetilde{C}_{-k-1}
  }{
    \widetilde{C}_{-k-1} - \widetilde{C}_{-k}
  }
  =
  \left(
    1- (1-\ell ) \, (1-m^2) \, \widetilde{C}_{-k-1}
  \right)
  \,
  \left(
    1- (1-\ell ) \, (1-m^2) \, \widetilde{C}_{-k-2}
  \right)
%
\end{equation}
with
\begin{gather*}
  \widetilde{C}_0(\ell,m)
  =
  \frac{1}{1+ \ell \, m^2}
  \\[2mm]
  \begin{aligned}
    & \widetilde{C}_{-1}(\ell,m)
    \\
    & =
    \frac{
      -\ell+ \ell \,  m^2 + m^4 + 2 \, \ell \, m^4
      + \ell^2 \,  m^4 + \ell \, m^6  - \ell \,  m^8
    }{
      -\ell+\ell^2 + 2 \, \ell \, m^2 - \ell^2 \, m^2
      + m^4 + 2 \, \ell \, m^4
      + 2 \, \ell^2 \, m^6 + \ell^3 \,  m^6 - \ell \,m^8
      + 2 \, \ell^2 \, m^8 + \ell \,m^{10} - \ell^2 \, m^{10}
    }
  \end{aligned}
\end{gather*}

\begin{prop}
  \label{prop:recursion_matrix}
  Rational function satisfying eq.~\eqref{recursion_tilde} is solved as
  \begin{equation}
    \label{define_A_B}
    \widetilde{C}_p(\ell,m)
    =
    \frac{
      A_p(\ell,m)
    }{
      B_p(\ell,m)
    }
  \end{equation}
  Here polynomials $A_p(\ell,m)$ and
  $B_p(\ell,m)$ are recursively defined as follows by use of a
  polynomial $Z(\ell,m)$ defined by
  \begin{equation}
    Z(\ell,m)
    =
    -\ell + \ell^2 + 2 \,  \ell \, m^2
    - \ell^2 \, m^2 + m^4 + \ell^2 \, m^4 - m^6
    + 2 \,  \ell \, m^6 + m^8 - \ell \,m^8
  \end{equation}
  \begin{itemize}
  \item
    a positive case, \emph{i.e.} twist knots $\mathcal{K}_{p>0}$
    \begin{align}
      \begin{pmatrix}
        A_{p+1}(\ell,m) \\[2mm]
        B_{p+1}(\ell,m)
      \end{pmatrix}
      & =
      \begin{pmatrix}
        -Z(\ell,m) & - (1-m^2) \, (\ell-m^4)
        \\[2mm]
         m^2 \, (\ell+m^2)^2 \, (1-\ell) \, (1-m^2) &
        - m^2 \, (\ell+m^2)^2
      \end{pmatrix}
      \,
      \begin{pmatrix}
        A_{p}(\ell,m) \\[2mm]
        B_{p}(\ell,m)
      \end{pmatrix}
      \nonumber      \\
      & \equiv
      \mathbf{M}_+
      \,
      \begin{pmatrix}
        A_{p}(\ell,m) \\[2mm]
        B_{p}(\ell,m)
      \end{pmatrix}
      \label{recursion_A_B}
    \end{align}
    with
    \begin{equation*}
      \begin{pmatrix}
        A_1(\ell,m) \\[2mm]
        B_1(\ell,m)
      \end{pmatrix}
      =
      \begin{pmatrix}
        \ell + m^6 \\[2mm]
        m^2 \, (\ell+m^2)^2 
      \end{pmatrix}
    \end{equation*}
  \item
    a negative case,
    \emph{i.e.}, twist knots $\mathcal{K}_{-p<0}$
    \begin{align}
      \begin{pmatrix}
        A_{-p-1}(\ell,m) \\[2mm]
        B_{-p-1}(\ell,m)
      \end{pmatrix}
      & =
      \begin{pmatrix}
        m^2 \, (\ell+m^2 )^2 & - (1-m^2) \, (\ell - m^4)
        \\[2mm]
        (1-\ell) \, (1-m^2) \, m^2 \, (\ell+m^2 )^2
        &
        Z(\ell,m)
      \end{pmatrix}
      \,
      \begin{pmatrix}
        A_{-p}(\ell,m) \\[2mm]
        B_{-p}(\ell,m)
      \end{pmatrix}
      \nonumber \\
      & \equiv
      \mathbf{M}_- \,
      \begin{pmatrix}
        A_{-p}(\ell,m) \\[2mm]
        B_{-p}(\ell,m)
      \end{pmatrix}
      \tag{\ref{recursion_A_B}$^\prime$}
      \label{recursion_A_B_negative}
    \end{align}
    with an initial condition
    \begin{equation*}
      \begin{pmatrix}
        A_0(\ell,m) \\[2mm]
        B_0(\ell,m)
      \end{pmatrix}
      =
      \begin{pmatrix}
        1 \\[2mm]
        1+ \ell \, m^2
      \end{pmatrix}
    \end{equation*}
  \end{itemize}
\end{prop}

We note that we have
\begin{gather}
  \left( \mathbf{M}_\pm \right)^t \cdot
  \boldsymbol{\sigma}_y \cdot
  \mathbf{M}_\pm \cdot
  \boldsymbol{\sigma}_y
  =
  m^4 \, (\ell+m^2 )^4
  \\[2mm]
  \mathbf{M}_+ \cdot  \mathbf{M}_-
  =
  -m^4 \, (\ell+m^2 )^4
\end{gather}
with the Pauli spin matrix
$\boldsymbol{\sigma}_y=
\begin{pmatrix}
  0 & - \mathrm{i} \\
  \mathrm{i} & 0
\end{pmatrix}
$,
and that
the characteristic polynomial of $\mathbf{M}_\pm$
is given by
\begin{multline}
  F_\pm(x)
  \\
  =
  x^2 \pm x
  \left(
    (\ell^2+m^4 ) \, (1+m^4 )
    +\ell \, (-1+2 \, m^2 + 2 \, m^4 + 2 \, m^6 - m^8)
  \right)
  +m^4 \, (\ell+m^2  )^4
\end{multline}

\begin{proof}[Proof of Prop.~\ref{prop:recursion_matrix}]
  We assume that
  $\widetilde{C}_p(\ell,m)$ is  defined by eq.~\eqref{define_A_B}, and
  that
  the polynomials
  $A_p$ and $B_p$ satisfy eqs.~\eqref{recursion_A_B}
  and~\eqref{recursion_A_B_negative}.
  Hereafter
  we use $A_p = A_p(\ell,m)$ and $B_p = B_p(\ell,m)$ for
  brevity.

  We first prove a positive case $p>0$.
  We see from eq.~\eqref{recursion_A_B} that
  \begin{gather*}
    (1-\ell) \, (1-m^2) \, A_p -  B_p
    =
    m^{-2} \, (\ell+m^2 )^{-2} \, B_{p+1}
    \\[2mm]
    A_p \, B_{p+1} - B_p \, A_{p+1}
    =
    m^4 \, (\ell+m^2 )^4 \,
    \left(
      A_{p-1} \, B_p - B_{p-1} \, A_p
    \right)
  \end{gather*}
  With these identities we find that
  \begin{multline*}
    \left(
      \frac{A_{p+1}}{B_{p+1}}
      -  \frac{A_{p+2}}{B_{p+2}}
    \right)
    \,
    \left(
      B_p -  (1-\ell) \, (1-m^2) \, A_p
    \right) \,
    \left(
      B_{p+1} -  (1-\ell) \, (1-m^2) \, A_{p+1}
    \right)
    \\
    =
    A_{p} \, B_{p+1} - B_{p} \, A_{p+1}
  \end{multline*}
  which proves eq.~\eqref{recursion_tilde}.

  For the negative case $-p<0$, 
  we easily see from a recursion relation~\eqref{recursion_A_B_negative} that
  \begin{gather*}
    B_{-p-1} - (1-\ell) \, (1-m^2) \, A_{-p-1}
    =
    m^2 \, (\ell+m^2)^2 \, B_{-p}
    \\[2mm]
    A_{-p-1} \, B_{-p-2} - B_{-p-1} \, A_{-p-2}
    =
    m^4 \, (\ell+m^2)^4 \,
    \left(
      A_{-p} \, B_{-p-1} - B_{-p} \, A_{-p-1}
    \right)
  \end{gather*}
  which leads
  \begin{multline*}
    \left(
      \frac{A_{-p}}{B_{-p}} -
      \frac{A_{-p-1}}{B_{-p-1}}
    \right)
    \left(
      B_{-p-1} - (1-\ell) \, (1-m^2) \, A_{-p-1}
    \right)
    \left(
      B_{-p-2} - (1-\ell) \, (1-m^2) \, A_{-p-2}
    \right)
    \\
    =
    A_{-p-1} \, B_{-p-2} - B_{-p-1} \, A_{-p-2}
  \end{multline*}
  This is nothing but eq.~\eqref{recursion_tilde_negative}
\end{proof}

\begin{prop}
  Polynomial $A_p(\ell,m)$
  defined by eq.~\eqref{define_A_B}
  coincides with  the A-polynomial for the twist knot
  $\mathcal{K}_p$.
\end{prop}

\begin{proof}
  We define a polynomial
  \begin{equation}
    X(\ell,m)
    =
    -\ell+ \ell^2 + 2 \, \ell \, m^2 + m^4 + 2 \, \ell \, m^4 +
    \ell^2 \, m^4 + 2 \, \ell \, m^6
    + m^8 - \ell \, m^8
  \end{equation}

  The recursion relation~\eqref{recursion_A_B} gives
  ($p>0$)
  \begin{gather}
    \label{recursion_A}
    A_{p+1}(\ell,m)
    =
    - X(\ell,m) \, A_{p}(\ell,m)
    - m^4 \,
    \left(
      \ell+m^2
    \right)^4 \, A_{p-1}(\ell,m)
  \end{gather}
  with
  \begin{align*}
    A_1(\ell,m)
    & =
    \ell+m^6
    \\
    A_2(\ell,m)
    & =
    \ell^2 - \ell^3 - 2 \, \ell^2 \, m^2 - \ell \,  m^4
    - 2 \,  \ell^2 \, m^4
    + \ell \, m^6 + \ell^2  \, m^8
    \\
    & \qquad \qquad
    - 2 \,  \ell \, m^{10}
    - \ell^2 \, m^{10} - 2 \, \ell \, m^{12} - m^{14} + \ell \, m^{14}
  \end{align*}
  For a negative case ($-p<0$)
  we also see that
  the recursion relation~\eqref{recursion_A_B_negative} reduces to
  \begin{gather}
    \label{recursion_A_negative}
    A_{-p-1}(\ell,m)
    = X(\ell,m) \, A_{-p}(\ell,m) 
    - m^4 \,
    \left(
      \ell+m^2
    \right)^4 \, A_{-p+1}(\ell,m)
  \end{gather}
  with 
  \begin{align*}
    A_0(\ell,m)&=1,
    \\
    A_{-1}(\ell,m)&=
    - \ell + \ell \, m^2
    + m^4 + 2 \, \ell \, m^4 + \ell^2 \, m^4 + \ell \, m^6
    - \ell \, m^8
  \end{align*}

  These recursion relations,
  eqs.~\eqref{recursion_A} and~\eqref{recursion_A_negative},
  coincide with those for the twist
  knot $\mathcal{K}_{p}$ derived in Ref.~\citen{HosteShana03a},
  and we can conclude that the polynomial $A_p(\ell,m)$ defined by
  eq.~\eqref{define_A_B} is the A-polynomial for  the twist knot
  $\mathcal{K}_p$.
\end{proof}

These propositions indicate
that
an algebraic equation~\eqref{f_and_1}, which is a consequence of
eqs.~\eqref{H_constraint} for the twist knot $\mathcal{K}_p$,
is nothing but an algebraic equation of the A-polynomial
for the twist knot $\mathcal{K}_p$;
\begin{equation}
  A_p(\ell, m)=0
\end{equation}
if we suppose $\ell \neq 1 $ and $m^2 \neq 1$.
As we define the parameter $\ell$, which now represents the eigenvalue of the
longitude of the boundary torus  of knot,
by a derivative of the function $H_\mathcal{K}(\boldsymbol{x},m^2)$ with respect to $m$,
we can identify the $H$-function with a constraint~\eqref{H_constraint_2} as
the potential function of the twist knot.
This proves  the statement of Theorem~\ref{prop:previous},
\emph{i.e.}, 
the $H$-function defined from
an asymptotics of  the colored Jones
polynomial~\eqref{Jones_and_integrand}
is the potential function under a constraint~\eqref{H_constraint_2},
and
it gives the A-polynomial
with a condition~\eqref{H_constraint_1}.
This fact may support the \emph{Volume Conjecture}~\cite{Kasha96b,MuraMura99a}
that the hyperbolic volume of the knot complements dominates an
asymptotics of the colored Jones polynomial,
as eq.~\eqref{H_constraint_2} denotes the saddle point equation of the
integral~\eqref{Jones_and_integrand}.
Indeed  we see that
under a constraint~\eqref{H_constraint_2} the $H$-function defined by
eqs.~\eqref{define_H} and~\eqref{define_H_negative} becomes
\begin{gather}
  \Imaginary H_{\mathcal{K}_{p}}(x_0, \dots, x_{|p|-1}, m^2=1)
  =
  3 \, D(1/x_0) +
  \sum_{i=0}^{|p|-1} D(x_i/x_{i+1})
\end{gather}
where
$(x_0,\dots,x_{|p|-1})$ is a solution of
eqs.~\eqref{saddle_positive_2}--\eqref{saddle_positive_3},
or
eqs.~\eqref{saddle_negative_2}--\eqref{saddle_negative_3},
under a constraint $m^2=1$.
Here we have used the Bloch--Wigner function $D(z)$ defined by
\begin{equation}
  D(z)
  =
  \Imaginary \Li(z) +
  \arg(1-z) \cdot \log |z|
\end{equation}
which denotes the hyperbolic volume of the ideal tetrahedron with
modulus $z$.

In the case of $m^2=1$
we can simplify those equations as follows.

\begin{prop}
  We consider the saddle point equations~\eqref{H_constraint_2},
  \emph{i.e.}
  eqs.~\eqref{saddle_positive_2}--\eqref{saddle_positive_3}
  or
  eqs.~\eqref{saddle_negative_2}--\eqref{saddle_negative_3},
  of the $H$-function in a case of $m^2 = 1$.
  Let the polynomial $V_k(z)$ be defined by ($k>0$)
  \begin{equation}
    \label{small_polynomial}    
    \begin{gathered}
      V_k(z)
      =
      \sum_{j=0}^{2 k}
      \begin{pmatrix}
        k + \left\lfloor \frac{j}{2} \right\rfloor \\
        j
      \end{pmatrix} \, z^j
      \\[2mm]
%
      V_{-k}(z)
      =
      1+ \sum_{j=1}^{2 k-1}
      \begin{pmatrix}
        k + \left\lfloor \frac{j-1}{2} \right\rfloor \\
        j
      \end{pmatrix} \, z^j
    \end{gathered}
  \end{equation}

  \begin{itemize}
  \item Positive case ($p>0$);
    
    Eqs.~\eqref{saddle_positive_2} and~\eqref{saddle_positive_3}
    with $m^2=1$
    are
    solved as
    \begin{equation}
      x_{p-k} = x_0 \, \frac{V_k(1-x_0)}{V_{k-1}(1-x_0)},
      \label{solution_m_zero}
    \end{equation}
    for $k=1,\dots,p$,
    and $x_0$ is a solution of
    $      V_p(1-x_0)=V_{p-1}(1- x_0)      $,
    \emph{i.e.},
    \begin{equation}
      \label{solution_small_x}
      V_{-p}(1-x_0)=0
    \end{equation}

  \item Negative case ($-p<0$);

    Eqs.~\eqref{saddle_negative_2} and~\eqref{saddle_negative_3}
    with $m^2=1$
    are
    solved as
    \begin{equation}
      x_{p-k} = x_0 \, \frac{V_{-k}(x_0-1)}{V_{-k-1}(x_0-1)},
      \tag{\ref{solution_m_zero}$^\prime$}
    \end{equation}
    for $k=1,\dots,p$,
    and $x_0$ is a solution of
    $V_{-p}(x_0-1)=V_{-p-1}(x_0-1)$,
    \emph{i.e.},
    \begin{equation}
      \tag{\ref{solution_small_x}$^\prime$}
      \label{solution_small_x_negative}
      V_p(x_0-1)=0
    \end{equation}
  \end{itemize}
\end{prop}
\begin{proof}
  We see that the polynomials $V_p(z)$ satisfy
  \begin{equation}
    \begin{gathered}
      V_p(z) - V_{p-1}(z) = z \, V_{-p}(z)
      \\[2mm]
      V_{-p-1}(z) - V_{-p}(z) = z \, V_{p}(z)
    \end{gathered}
  \end{equation}
  which gives the 3-term relations
  \begin{equation}
    \label{recursion_polynomial}
    \begin{gathered}
    V_{p+1}(z) - \left( z^2 +2 \right) \, V_p(z) + V_{p-1}(z) = 0
    \\[2mm]
    V_{-p-1}(z) - \left( z^2 +2 \right) \, V_{-p}(z) + V_{-p+1}(z) = 0
    \end{gathered}
  \end{equation}

  To complete the proof,
  we need to  show that the solution of eqs.~\eqref{define_C}
  and~\eqref{define_C_negative} with $m^2=1$ is given by
  $\frac{V_k(1-x_0)}{V_{k-1}(1-x_0)}$ and
  $\frac{V_{-k}(x_0-1)}{V_{-k-1}(x_0-1)}$ respectively,
  \emph{i.e.},
  the polynomial satisfies the bilinear equation
  \begin{align*}
    V_{k+2}(z) \cdot V_k(z) - \bigl( V_{k+1}(z) \bigr)^2
    & =
    V_{k+1}(z) \cdot V_{k-1}(z) - \bigl( V_{k}(z) \bigr)^2
    \\
    & = -z^3 
    \\[2mm]
    V_{-k-2}(z) \cdot V_{-k}(z) - \bigl( V_{-k-1}(z) \bigr)^2
    & =
    V_{-k-1}(z) \cdot V_{-k+1}(z) - \bigl( V_{-k}(z) \bigr)^2
    \\
    & = z^3 
  \end{align*}
  This can be done easily by induction using
  eq.~\eqref{recursion_polynomial}.
\end{proof}

We note that the polynomials $V_k(z)$ are  written as a sum of the
hypergeometric functions,
\begin{equation}
  \begin{gathered}
    V_k(z)
    =
    {}_2 F_1
    \left(
      \begin{array}{c}
        -k, k+1
        \\
        \frac{1}{2}
      \end{array}
      ; - \frac{z^2}{4}
    \right)
    +
    k \, z \cdot
    {}_2 F_1
    \left(
      \begin{array}{c}
        1-k, k+1
        \\
        \frac{3}{2}
      \end{array}
      ; - \frac{z^2}{4}
    \right)
    \\[2mm]
    V_{-k}(z)
    =
    {}_2 F_1
    \left(
      \begin{array}{c}
        1-k, k
        \\
        \frac{1}{2}
      \end{array}
      ; - \frac{z^2}{4}
    \right)
    +
    k \, z \cdot
    {}_2 F_1
    \left(
      \begin{array}{c}
        1-k, k+1
        \\
        \frac{3}{2}
      \end{array}
      ; - \frac{z^2}{4}
    \right)
  \end{gathered}
\end{equation}

With these results
we  conclude that
when $x_0$ is a solution of eq.~\eqref{solution_small_x}
or~\eqref{solution_small_x_negative}
we have ($p>0$)
\begin{multline}
  \Imaginary H_{\mathcal{K}_{p}}(x_0, \dots, x_{p-1}, m^2=1)
  \\
  =
  3 \, D(1/x_0) +
  \sum_{j=1}^{p-1}
  D\left(
    \frac{V_{j+1}(1-x_0) \, V_{j-1}(1-x_0)}{
      \left(  V_j(1-x_0)  \right)^2
    }
  \right)
  +
  D \left( x_0 \, \frac{V_1(1-x_0)}{V_0(1-x_0)} \right)
  \label{volume_x0}
\end{multline}
and
\begin{multline}
  \tag{\ref{volume_x0}$^\prime$}
  \label{volume_x0_negative}
  \Imaginary H_{\mathcal{K}_{-p}}(x_0, \dots, x_{p-1}, m^2=1)
  \\
  =
  3 \, D(1/x_0) +
  \sum_{j=1}^{p}
  D\left(
    \frac    {
      \left(   V_{-j}(x_0-1)  \right)^2
    }{
      V_{-j+1}(x_0-1) \, V_{-j-1}(x_0-1)
    }
  \right)
\end{multline}

\begin{table}[htbp]
  \centering
  \begin{equation*}
    \def\arraystretch{1.3}
    \begin{array}{c||c|c}
      p &
      \left.
        \Imaginary H_{\mathcal{K}_p}
      \right|_{m^2=1}
      =
      \Vol(S^3 \setminus \mathcal{K}_p)
      &
      x_0
      \\
      \hline \hline
      -5 &
      3.57388
      &
      0.99151 -
      1.91177 \, \mathrm{i}
      \\
      \hline
      -4 &
      3.52620 
      &
      0.98405 -
      1.86641 \, \mathrm{i}
      \\
      \hline
      -3  &
      3.42721
      &
      0.96453 -
      1.77530 \, \mathrm{i}
      \\
      \hline
      -2 &
      3.16396 
      &
      0.89512 -
      1.55249 \, \mathrm{i}
      \\
      \hline
      -1 &
      2.02988
      &
      0.50000 -
      0.86603 \,\mathrm{i}
      \\
      \hline
      2 &
      2.82812
      &
      1.21508 -
      1.30714 \, \mathrm{i}
      \\
      \hline
      3 &
      3.33174 
      &
      1.05818-
      1.69128 \, \mathrm{i}
      \\
      \hline
      4 &
      3.48666 
      &
      1.02317 -
      1.82953 \, \mathrm{i}
      \\
      \hline
      5 &
      3.55382
      &
      1.01144 -
      1.89257 \, \mathrm{i}
      \\
      \hline
    \end{array}
  \end{equation*}
  \caption{Hyperbolic volume of the complement of the twist knot
    $\mathcal{K}_p$
    coincides with the largest  value of $\Imaginary H_{\mathcal{K}_p}$.
    Given are values of $x_0$ which give
    the hyperbolic    volume $\Vol(S^3 \setminus \mathcal{K})$
    by eq.~\eqref{volume_x0} or~\eqref{volume_x0_negative}.
    Knot $\mathcal{K}_{p=1}$ is the left-hand
    trefoil, which is not hyperbolic.
  }
  \label{tab:volume}
\end{table}

\begin{figure}[htbp]
  \centering
  \psfrag{a}{$\Real z$}
  \psfrag{b}{$\Imaginary z$}
  \includegraphics[scale=1.2]{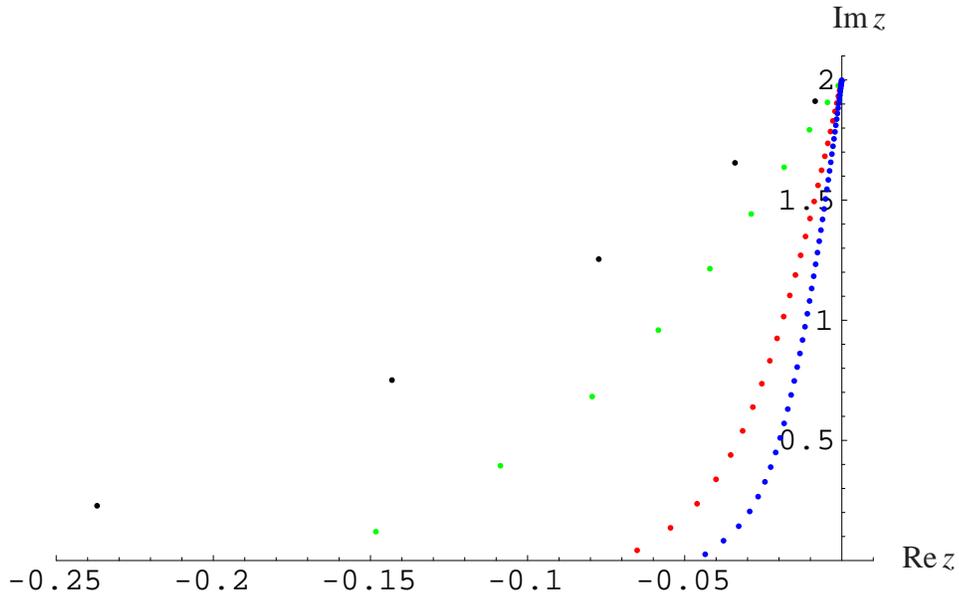}
  \caption{Zeros of the polynomials $V_k(z)$ for
    $k=5 ({\color{black} \bullet}),
    10 ({\color{green} \bullet}),
    30 ({\color{red} \bullet}),
    50 ({\color{blue} \bullet})$.
    We only plot zero points in the upper half plane.
  }
  \label{fig:zero}
\end{figure}

See Table~\ref{tab:volume} for numerical computation.
We have checked  that the  largest 
value of $\Imaginary H_{\mathcal{K}_p}(x_0,\dots,x_{|p|-1}, m^2=1)$
among solutions of eq.~\eqref{H_constraint_2}
coincides with the hyperbolic volume of the complement of
$\mathcal{K}_p$~\cite{SnapPea99}
as was proposed as \emph{Volume Conjecture}
(see  Ref.~\citen{SakumWeek95a} for an ideal triangulation of the
complement of the twist knots).
We have plotted zeros of the polynomial $V_k(z)$ 
in
Fig.~\ref{fig:zero} for convention.

\begin{figure}[htbp]
  \centering
  \includegraphics[scale=0.57]{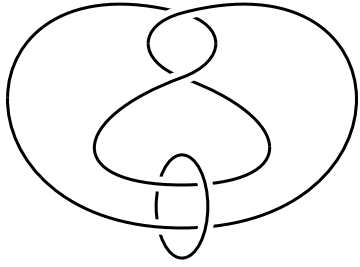}
  \caption{Whitehead link}
  \label{fig:Whitehead}
\end{figure}

In a limit $p\to\infty$, we may read off from both
the above table and a
numerical computation
that $x_0\to 1 - 2 \, \mathrm{i}$
(see Fig.~\ref{fig:zero}).
It is known  geometrically that  in the
limit of  $|p|\to\infty$
the hyperbolic volume of the twist knot $\mathcal{K}_p$ is
that of the Whitehead link~\cite{JMiln82a}
(Fig.~\ref{fig:Whitehead}),
which coincides with the hyperbolic volume of the regular ideal
octahedron
$4 \, D(\mathrm{i}) = 3.66386237670887606 \dots$.
Applying identities
\begin{equation}
  \begin{aligned}
    V_p(2 \, \mathrm{i})
    & = (-1)^p \, \left(
      2 \, p + 1 -2 \,   p  \, \mathrm{i} 
    \right)
    \\[2mm]
    V_{-p}(-2 \, \mathrm{i})
    & = (-1)^p \, \left(
      -2 \, p + 1 + 2 \,   p  \, \mathrm{i} 
    \right)
  \end{aligned}
\end{equation}
for $p >0$, which result from the Chu--Vandermonde identity,
to eqs.~\eqref{volume_x0} and~\eqref{volume_x0_negative},
we may obtain  formulae for the Bloch--Wigner function;
  \begin{align}
    4 \, D(\mathrm{i})
    & =
    3 \, D(2 \, \mathrm{i})
    +
    \sum_{k=0}^\infty
    D
    \left(
      \left(
        k+\frac{1}{4} +\frac{\mathrm{i}}{4}
      \right)^2
    \right)
    \label{sum_D_function}
    \\
    & =
    3 \, D(2 \, \mathrm{i})
    -
    \sum_{k=0}^\infty
    D \left(
      \left(
        -k -\frac{3}{4} 
        +        \frac{\mathrm{i}}{4}
      \right)^2
    \right)
    \tag{\ref{sum_D_function}$^\prime$}
  \end{align}
In fact these identities
follow from
the pentagon identity,
especially  an identity
$D(z^2)=2 \,\left(  D(z) - D(z+1) \right)$
~\footnote{
  Anatol N. Kirillov kindly pointed out this fact.}.

\section{Torus Knot}
\label{sec:torus}

\begin{figure}[htbp]
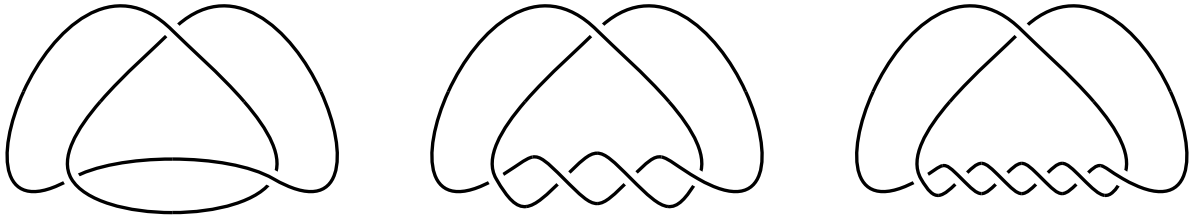

  \centering
  \begin{tabular}{cp{5mm}cp{5mm}c}
    \includegraphics[scale=0.63]{torus23.ps}
    & &
    \includegraphics[scale=0.63]{torus52.ps}
    & &
    \includegraphics[scale=0.63]{torus72.ps}
  \end{tabular}
  \caption{Torus knots $\mathcal{T}_{2,2p+1}$ with $p>0$
    are depicted.
    {}From left to right,
    $p=1$ (right-hand trefoil),
    $p=2$ (Solomon's seal knot), and $p=3$.}
  \label{fig:torus}
\end{figure}

We apply above story to the colored Jones
polynomial  for the torus
knot $\mathcal{T}_{2,2  p+1}$
(we study a case of $p>0$. See Figure~\ref{fig:torus});
we reveal a relationship between the A-polynomial and the $H$-function.

We first recall the $q$-hypergeometric expression of the colored Jones
polynomial for the torus knot.
\begin{prop}[\cite{KHikami04a}]
  The $N$-colored Jones polynomial
  $J_{\mathcal{K}}(N)$ for the torus knot
  $\mathcal{K}=\mathcal{T}_{2,2p+1}$
  is written as
  \begin{multline}
    \label{Jones_torus}
    J_{\mathcal{T}_{2,2  p+1}}(N)
    =
    q^{ p \, (1-N^2)}
    \sum_{s_p \geq \dots \geq s_2 \geq s_1 \geq 0}^\infty
    q^{(p-1) \, s_p  \, (s_p+1) } \,
    \frac{
      (q^{1-N})_{s_p} \, (q^{1+N})_{s_p}
    }{
      (q)_{s_p}
    }
    \\
    \times
    \left(
      \prod_{i=1}^{p-1}
      q^{s_i^{~2}-(2 s_p +1) s_i} \,
      \begin{bmatrix}
        s_{i+1} \\
        s_i
      \end{bmatrix}_q
    \right) \,
  \end{multline}
  Here the colored Jones polynomial is normalized to be
  $J_{\text{unknot}}(N)=1$.
\end{prop}


Applying eq.~\eqref{log_and_dilog} to above expression,
we easily obtain the $H$-function
for the torus knot.

\begin{prop}
  The asymptotic behavior of the  $N$-colored Jones polynomial for the torus knot
  $\mathcal{T}_{2,2p+1}$
  is written in an integral form
  in a limit $N\to\infty$
  as
  \begin{equation}
    J_{\mathcal{T}_{2,2 p+1}}(N)
    \sim
    \iiint \mathrm{d} x_0 \cdots \mathrm{d} x_{p-1}
    \exp
    \left(
      \frac{N}{2 \, \pi \, \mathrm{i} \, r} \,
      H_{ \mathcal{T}_{2,2 p+1}}(x_0, \dots, x_{p-1}, m^2 )
    \right)
  \end{equation}
  where $m$ is defined by eq.~\eqref{define_m}, and
  \begin{multline}
    H_{ \mathcal{T}_{2,2 p+1}}(x_0, \dots, x_{p-1}, m^2 )
    =
    - p \, \left( \log ( m^2 )\right)^2
    +
    \sum_{i=1}^{p-1}
    \left(
      \log \left({x_i}/{x_0}\right)
    \right)^2
    \\
    +
    \Li ( m^2 ) + \Li(1/m^2)
    - \Li(x_0/m^2) - \Li(m^2 \, x_0)
    +
    \sum_{i=0}^{p-1} \Li(x_{i}/x_{i+1}) - p \, \frac{\pi^2}{6}
    \label{torus_define_H}
  \end{multline}
\end{prop}

The  theorem for the torus knots is as follows.

\begin{theorem}
  The $H$-function~\eqref{torus_define_H}
  is the potential function for the torus knot $\mathcal{T}_{2,2p+1}$
  under a constraint~\eqref{H_constraint_2},
  and it
  gives the A-polynomial  for the torus knot $\mathcal{T}_{2,2p+1}$
  by eliminating $\boldsymbol{x}$ with a help of
  a condition~\eqref{H_constraint_1}.
\end{theorem}

\begin{proof}
  A set of equation~\eqref{H_constraint} gives
  \begin{subequations}
    \label{torus_constraint}
    \begin{gather}
      \frac{1-m^2 \, x_0}{
        m^{4 p} \,
        \left(
          x_0 - m^2
        \right)
      } = \ell,
      \\[2mm]
      x_0^{~2 p-2} \,
      \left(
        1- m^2 \, x_0
      \right) \,
      \left(
        1- \frac{x_0}{m^2}
      \right)
      =
      \left(
        1-\frac{x_0}{x_1}
      \right) \,
      \left(
        \prod_{i=1}^{p-1} x_i^{~2}
      \right)
      \\[2mm]
      x_i^{~2} \,
      \left(
        1- \frac{x_{i-1}}{x_i}
      \right)
      =
      x_0^{~2} \,
      \left(
        1- \frac{x_i}{x_{i+1}}
      \right)
      \qquad
      \text{for $i=1,2,\dots,p-1$}
    \end{gather}
  \end{subequations}
  In this case we have
  \begin{equation}
    x_0 =
    \frac{1 + \ell \, m^{4 p +2}}{
      m^2 \, 
      \left(
        1 + \ell \, m^{4 p-2}
      \right)
    }
    \label{torus_x_0}
  \end{equation}
  and we see that
  \begin{equation}
    x_k= x_0 \, C_{p-k}(x_0)
  \end{equation}
  where $C_k$ is recursively solved as
  \begin{gather}
    C_{k+1}(x)
    =
    \left(
      1-\frac{(1-m^2 \, x) \, (m^2 -x )}{
        m^2 \, \left( C_1(x) \cdots C_{k}(x) \right)^2
      }
    \right)
    \, C_k(x)
    \\[2mm]
    C_0(x)=\frac{1}{x}
    \nonumber
  \end{gather}
  Then the algebraic equation for $\ell$ and $m$  may reduce to
  \begin{equation}
    C_p(x_0)
    = 1
  \end{equation}
  where $x_0$ is solved in eq.~\eqref{torus_x_0}.
  In this case we have
  \begin{equation*}
    1-C_p(x_0) = \frac{(1-\ell) \, (1-m^2)}{1 + \ell \, m^2}
  \end{equation*}
  and we  obtain an unwanted  solution $\ell =1$ or $m^2 =1$.
  This suggests that a solution of eqs.~\eqref{torus_constraint} is
  rather given by
  \begin{align*}
    x_0 & = 0,
    &
    x_i & = \pm 1,
    \quad \text{for $i>0$}
  \end{align*}
  This  gives
  an algebraic equation as
  \begin{equation}
    A_{\mathcal{T}_{2,2p+1}}(\ell,m)
    = 
    1+ \ell \, m^{4 p+2}
  \end{equation}
  which is the A-polynomial for the torus knot
  $\mathcal{T}_{2,2 p+1}$~\cite{CCGLS94a,HosteShana03a}.
\end{proof}

We note that an exact asymptotic expansion in $N\to\infty$
of the $N$-colored Jones polynomial for the torus knot $\mathcal{T}_{s,t}$
with $q$ being the $N$-th root of unity,
$q=\exp(2 \, \pi \, \mathrm{i}/N)$,
was studied in
Refs.~\citen{DZagie01a,KHikami02c,KHikami03c}
(see also Refs.~\citen{KHikami03a,KHikami04b}),
and  the invariant was identified with the Eichler integral of the
modular form with   half-integral weight $1/2$
which is related to
the character of the Virasoro minimal model $\mathcal{M}(s,t)$.

We shall restate
our theorems for the  trefoil $\mathcal{T}_{2,3}$ in more detail.
The  $N$-colored Jones polynomial for the right-hand  trefoil
was  computed explicitly
also in Refs.~\citen{TQLe03a,KHabi02a},
and
collecting these results
we have
\begin{subequations}
  \label{Jones_trefoil}
  \begin{align}
    J_{\mathcal{T}_{2,3}}(N)
    & =
    q^{1-N} \sum_{n=0}^\infty
    q^{-n \, N} \, (q^{1-N})_n
    \\
    & =
    \sum_{n=0}^\infty q^{-n (n+2)} \,
    (q^{1-N})_n \, (q^{1+N})_n
    \\
    & =
    q^{1-N^2}
    \sum_{n=0}^\infty
    \frac{(q^{1-N})_n \, (q^{1+N})_n}{
      (q)_n
    }
  \end{align}
\end{subequations}
All these infinite sums reduce into  finite sums due to $(q^{1-N})_k=0$ for
$k \geq N >0$.
Note that those $q$-hypergeometric type expressions
are respectively from  Refs.~\citen{TQLe03a,KHikami04a},
eq.~\eqref{Jones_twist} with $p=1$
replacing $q$ by $q^{-1}$, and eq.~\eqref{Jones_torus} with $p=1$.

By use of eq.~\eqref{log_and_dilog}
we obtain the  $H$-functions from three expressions~\eqref{Jones_trefoil}
as follows;
\begin{subequations}
  \begin{align}
    H_a(x, m^2)
    & =
    -  ( \log  x ) \, \left( \log (m^2) \right)
    + \Li \Bigl(\frac{1}{m^2}\Bigr) - \Li\Bigl(\frac{x}{m^2}\Bigr)
    \label{H_trefoil_a}
    \\[2mm]
    H_b(x,m^2)
    & =
    - \left( \log x \right)^2
    + \Li\Bigl(\frac{1}{m^2}\Bigr) + \Li(m^2)
    - \Li\Bigl(\frac{x}{m^2}\Bigr) - \Li(m^2 \, x) 
    \label{H_trefoil_b}
    \\[2mm]
    H_c(x, m^2)
    & =
    -\left( \log (m^2) \right)^2
    + \Li\Bigl(\frac{1}{m^2}\Bigr) + \Li(m^2)
    - \Li\Bigl(\frac{x}{m^2}\Bigr) - \Li(m^2 \, x) + \Li(x) -
    \frac{\pi^2}{6}
    \label{H_trefoil_c}
  \end{align}
\end{subequations}
A set of equations~\eqref{H_constraint} is solved as follows;
\begin{enumerate}
  \def\labelenumi{(\alph{enumi})}
\item We have from eq.~\eqref{H_trefoil_a}
  \begin{align*}
    & \frac{-1+m^2}{(m^2-x) \, x} = \ell
    &
    & \frac{m^2-x}{m^4}=1
  \end{align*}
  from which we have $x=(1-m^2) \, m^2$.
  We thus  obtain an algebraic equation, $A(\ell,m)=0$,
  with
  \begin{equation}
    \label{A_trefoil}
    A(\ell,m) = 1 + \ell  \, m^6
  \end{equation}
  This is the A-polynomial for the
  (right-hand)
  trefoil~\cite{CCGLS94a}.

\item
  Substituting eq.~\eqref{H_trefoil_b}  for
  eqs.~\eqref{H_constraint}, we get
  \begin{align*}
    & \frac{-1+ m^2 \, x}{m^2-x} = \ell
    &
    & \frac{(m^2-x) \, (1-m^2 \, x)}{m^2 \, x^2}=1
  \end{align*}
  This gives  $x=\frac{1+\ell \, m^2}{m^2 + \ell}$,
  and
  an equation of $\ell$ and $m$ is written as
  \begin{equation*}
    \frac{(\ell+m^2) \, (1+\ell \, m^6)}{m^2 \, (1+\ell \, m^2)^3}
    =0
  \end{equation*}
  which suggests eq.~\eqref{A_trefoil}.

\item We have
  \begin{align*}
    & \frac{1-m^2 \, x}{m^4 \, (m^2-x)}=\ell
    &
    & \frac{(m^2-x) \, (1- m^2 \, x)}{m^2 \, (1-x)}=1
  \end{align*}
  which gives $x=\frac{1+ \ell \, m^6}{m^2 \, (1 + \ell \, m^2)}$
  and
  \begin{equation*}
    \frac{(1-\ell) \, (1+ \ell \, m^6)}{
      (1 + \ell \, m^2) \, ( 1 - \ell \, m^4)
    }=0
  \end{equation*}
  We may assume $\ell \neq 1$, and
  we obtain the A-polynomial~\eqref{A_trefoil}.
\end{enumerate}
To conclude,
all three $H$-functions given from an asymptotics of  three expressions~\eqref{Jones_trefoil},
give the A-polynomial for the trefoil with constraints~\eqref{H_constraint}.

\section*{Acknowledgments}
The author would like to thank
J.~Kaneko,
A.~N.~Kirillov,
H.~Murakami,
T.~Takata,
and Y.~Yokota
for discussions.
This work is supported in part by Grant-in-Aid for Young Scientists
from the Ministry of Education, Culture, Sports, Science and
Technology of Japan.


\end{document}